\documentstyle[12pt]{article}
\title{Two-band model and MgB$_2$ superconductivity}
\author{N.Kristoffel$\sp{1}$\thanks{e-mail: kolja@fi.tartu.ee}
and T.\"Ord$\sp 2$\\
$\sp 1$Institute of Physics, University of Tartu,\\
Riia 142, 51014 Tartu, Estonia\\
$\sp 2$Institute of Theoretical Physics, University of Tartu\\
T\"ahe 4, 51010 Tartu, Estonia}
\date{}
\begin{document}
\maketitle
\noindent\underline{Abstract.} A simple two-band model with interband
scattering of intraband pairs is applied to MgB$_2$ superconductivity.
An adjustable parameters set is chosen. The calculated $T_c$ vs doping,
isotope exponent and the heat capacity jump are in qualitative
agreement with the observations.\\


The discovery [1] of MgB$_2$ superconductivity with $T_c$=39 K has
qualitatively enlarged the family of high-temperature superconductors
and stimulated rapid research activity.

When asking about the pairing mechanism in this compound the presence
of boron isotope effect [2], the s-wave nature of the order parameter(s)
[3-5,17] and the hole-type conductivity [6,12] are of guiding significance.
Doping of MgB$_2$ with electrons (Mg$_{1-x}$Al$_x$B$_2$) reduces $T_c$
[7]. The same is observed for the (enhanced) hole doping
(Mg$_{1-x}$Li$_x$B$_2$) [8]. Considering the whole class of doped
compounds as in cuprates with a bell-type curve of $T_c(x)$ one expects
MgB$_2$ being positioned nearly in the optimal doping region of the
phase diagram.

High phonon frequencies by the small boron mass and strong electron-phonon
interaction have stimulated the description of the MgB$_2$ superconductivity
on a BCS type way [9,10]. However seemingly a limiting ability of such
a type mechanism is exploited at this. An original approach to MgB$_2$
superconductivity has been elaborated [11,12] using a "hole-undressing''
theory [13].

The electronic structure of MgB$_2$ is known by a number of
calculations [9,10,14]. The most impressive circumstance consists
here in the presence of two close genealogically boron $p_{x,y}$ bands
intersecting the Fermi surface. These $\sigma$-bonding hole-conducting
two-dimensional bands can be modulated by $E_{2g}$-boron stretching in
plane modes with a large deformation potential.

Presence of two bands near the Fermi energy in the electron spectrum
of cuprates has stimulated the application of various models with
interband interactions to explain their superconductivity, for a review
e.g. [15]. The physical origin of these bands has remained often unclear,
however a number of provising results have been obtained. This experience
stimulates the attempt to use the two-band model also to describe the
MgB$_2$ superconductivity which is the aim of the present note (see
also [16]). References to the nature of the MgB$_2$ electron spectrum
and to the specific heat data [17] serving evidence of a multicomponent
gap can be made at this.

We describe a two-band s-wave superconductor with the interband
scattering of intraband pairs by the mean-field Hamiltonian
\begin{eqnarray}
H & = & \sum_{\vec{k}\sigma}[\epsilon_a(\vec{k})a_{\vec{k}\sigma}\sp +
a_{\vec{k}\sigma}+ \epsilon_b(\vec{k})b_{\vec{k}\sigma}\sp +
b_{\vec{k}\sigma}] \nonumber\\
& + & \Delta_a\sum_{\vec{k}} [a_{\vec{k}\uparrow}a_{-\vec{k}\downarrow}+
a_{-\vec{k}\downarrow}\sp +a_{\vec{k}\uparrow}\sp +]-
\Delta_b\sum_{\vec{k}} [b_{\vec{k}\uparrow}b_{-\vec{k}\downarrow}+
b_{-\vec{k}\downarrow}\sp +b_{\vec{k}\uparrow}\sp +]\; ,
\end{eqnarray}
where
\begin{equation}
\Delta_a=2W\sum_{\vec{k}}\langle b_{\vec{k}\uparrow}
b_{-\vec{k}\downarrow}\rangle\; ,
\quad \Delta_b=2W\sum_{\vec{k}}
\langle a_{-\vec{k}\downarrow}a_{\vec{k}\uparrow} \rangle \; .
\end{equation}

Usual designations of electron operators and spins are used. The band
energies read $\xi_{a,b}(\vec{k})=\epsilon_{a,b}(\vec{k})+\mu$, $\mu$ is the
chemical potential. The interband interaction constant $W$ is supposed
to contain an electronic (Coulomb) part $U$ besides an electron-phonon
contribution $V\sim M\sp{-1}$ (both repulsive). Interband type pairs
interaction is omitted at present with the usual argument of much smaller
momentum volume available. The gap equations ($\Theta =k_BT$)
\begin{eqnarray}
\Delta_a & = & W\Delta_b\sum_{\vec{k}}E_b\sp{-1}(\vec{k})th
\frac{E_b(\vec{k})}{2\Theta} \\
\Delta_b & = & W\Delta_a\sum_{\vec{k}}E_a\sp{-1}(\vec{k})th
\frac{E_a(\vec{k})}{2\Theta} \nonumber
\end{eqnarray}
with quasiparticle energies $E_{a,b}=\sqrt{\epsilon_{a,b}\sp 2(\vec{k})+
\Delta_{a,b}\sp 2}$ determine the transition temperature $\Theta_c$
when $\Delta_{a,b}$ simultaneously tend to zero.

The MgB$_2$ electron spectrum in the region of interest can be
modelled by two plane parabolic hole bands with the densities of states
$\rho_a=0.18$, $\rho_b=0.07 (eV)\sp {-1}$ [10]. Their common origin is taken as
energy zero and the chemical potential of the undoped material  is
$\mu=-0.6$ eV. Looking on the common region of the bands according to
calculations [9,10] we introduce cut-off of the interband interaction
at $D=-1.5$ eV.

Performing the approximate integration in the equation
\begin{equation}
1 = W\sp 2 \rho_a\rho_b[\int_D\sp 0 \frac{d\xi}{\xi -\mu}
th\frac{\xi - \mu}{2\Theta_c}]\sp 2
\end{equation}
for $\Theta_c$, one finds ($ln\gamma =0.577$)
\begin{equation}
\Theta_c=\frac{2\gamma}{\pi}\sqrt{\mu (D -\mu)}
exp \{ -\frac{1}{2W\sqrt{\rho_a\rho_b}} \} \; .
\end{equation}

This expression reflects the usual advantage of two-band models to
work with repulsive interband interaction $W>0$ (note opposite signs of
order parameters at this [18]) and electronic energy scale in the
prefactor of the exponent. The choice of a common $D$ for the bands
leads to symmetric $\Theta_c(\mu )$ curve. For the characteristic
$\frac{2\Delta_{a,b}}{\Theta_c}$ ratios the BCS universality breaking
is expected. For the isotope effect exponent the simple result
independent of $\mu$ follows [21]
\begin{equation}
\alpha = \frac{C_B}{2W\sqrt{\rho_a\rho_b}}\frac{V}{W}\; ,
\end{equation}
where $C_B=dln M/d ln M_B$ ($M_B$ is the mass of boron ion) determines the participation of the boron
atoms in the vibration serving the interband electron-phonon
contribution $V\sim M\sp{-1}$. From (6) it is seen that factors which
enhance $T_c$ reduce its isotope effect exponent. Relatively small
electron-phonon contribution to the whole $W$ is able to cause a
remarkable isotope effect.

Hydrostatic pressure must shorten also the interplane B-B distances
and splits the actual bands [10]. This must lead according to the present
scheme to $T_c$ reduction as observed [19]. Then also the underdoped
region will be enlarged by the $\mu$ out of the bands overlap
situation. A pseudogap type excitation channel connected with the
lower band can be expected in this case as in cuprates [20].

For the specific heat jump at $T_c$ on finds using [15]
\begin{equation}
\Delta C = 9.42\; Rk_B\Theta_c \; , \; \; \; R = \rho_a\gamma_a\sp 2+
\rho_b\gamma_b\sp 2\; ,
\end{equation}
$$
\gamma_a=\left\{ \frac{1+W\sp 2\rho_a\rho_bA}
{1+W\sp 4\rho_a\sp 3\rho_bA\sp 2}\right\}\sp{1/2}\; , \; \; \; \;
\gamma_b=\left\{ \frac{1+W\sp 2\rho_a\rho_bA}
{1+W\sp 4\rho_a\rho_b\sp 3A\sp 2}\right\}\sp{1/2}
$$
$$
A = ln\sp 2 [\mu (D-\mu )4\gamma\sp 2 (\pi\Theta_c)\sp{-2}]\; .
$$

In quantitative estimations we use the value $W=0.81$ eV to reach the
measured $\Theta_c=40$ K for $\mu =-0.6$ eV. The ionic type
Mg$\sp{2+}$(B$\sp -$)$_2$ configutaion can play a role in determining
this scale of $W$. The observed $\alpha =0.26$ [2] determines for
$C_b=1$ (only boron atoms vibrate) the relation $V/W=4.7\%$ which
compares with the result found for cuprates [21,22].

The $\Theta_c(x)$ curve is given in Fig.1. As seen, MgB$_2$ can be
considered as belonging to optimally (auto-)doped region. A
further hole doping must soon (after passing the optimal region) lead
to $\Theta_c$ decrease as also the electron doping in agreement
with the experiment [7,8]. The heat capacity jump vs doping curve is
given in Fig.2. It resembles the $\Theta_c$ behaviour. Our result for
MgB$_2$ $\Delta C\sim 50$ $\frac{{\rm mJ}}{{\rm mol K}}$ can be
compared with the measured [23] value $\sim 86$
$\frac{{\rm mJ}}{{\rm mol K}}$, and $\Delta C/T_c = 1.3$
$\frac{{\rm mJ}}{{\rm mol K\sp 2}}$ with the experimental result
2 $\frac{{\rm mJ}}{{\rm mol K\sp 2}}$. The heat capacity characteristics
do not contain further free parameters and serve as a proof of the
applicability of the present theoretical scheme with the parameters set used.

A slight increase of $\rho_a\rho_b$ can significantly improve the
agreement with the measured heat capacity data without notable
modifications of other aspects. We note also that $\rho_{a,b}$ values
appropriate for the full band widths (5.6 and 14 eV [10]) have been
used and the boron $p_z$-type band  has remained out of play.

We find that multiband-type models remain of interest for the
explanation of doped MgB$_2$-family superconductivity.

This work has been supported by Estonian Science Foundation
Grant No 3591.

\subsection*{References}

\begin{enumerate}
\item J.Nagamatsu, N.Nakagawa, T.Muranaka, Y.Zenitani and J.Akimitsu,
Nature {\bf 410}, 63 (2001).
\item S.L.Bud'ko et al., Phys. Rev. Lett. {\bf 86}, 1877 (2001).
\item G.Karapetrov et al., cond.-mat./ 0102312 (2001).
\item H.Schmidt et al., cond.-mat./ 0102389 (2001).
\item H.Kotegawa et al., cond.-mat./ 0102334 (2001).
\item W.N.Kang et al., cond.-mat./ 0102313 (2001).
\item Y.Slusky et al., cond.-mat./0102262 (2001).
\item Y.G.Zhao et al., cond.-mat./0103077 (2001).
\item J.Kortus et al., cond.-mat./0101446 (2001).
\item J.M.An, W.E.Pickett, cond.-mat./0102391 (2001).
\item J.E.Hirsh and F.Marsiglio, cond.-mat./0102479 (2001).
\item J.E.Hirsch, con.-mat./0102115 (2001).
\item J.E.Hirsch, Phys. Rev. B, {\bf 62}, 14487 (2001).
\item K.D.Belashchenko et al., cond.-mat./0102290 (2001).
\item N.Kristoffel, P.Konsin and T.\"Ord, Rivista Nuovo Cim., {\bf 17}, No 9, 1 (1994).
\item M.Imada, cond.-mat./0103006 (2001).
\item Y.Wang, T.Plackowski and A.Junod, cond.-mat./0103181 (2001).
\item N.Kristoffel, T.\"Ord and P.Konsin, Nuovo Cim. {\bf 16D}, 311 (1994).
\item B.Lorenz, R.L.Meng and C.W.Chu, cond.-mat./0102264 (2001).
\item N.Kristoffel and P.Rubin, Physica C (to be published).
\item P.Konsin, N.Kristoffel, T.\"Ord, Ann. Physik {\bf 2}, 279 (1993).
\item T.\"Ord and N.Kristoffel, J. Low Temp. Phys. {\bf 117}, 253 (1999).
\item R.K.Kremer, B.J.Gibson and K.Ahn, cond.-mat./0102432 (2001).
\end{enumerate}

\newpage

Figure captions

\vspace{2cm}

\noindent Fig. 1. Dependence of $T_c$ on chemical potential position.

\vspace{1cm}

\noindent Fig. 2. Dependence of specific heat jump at $T=T_c$ on chemical potential position.

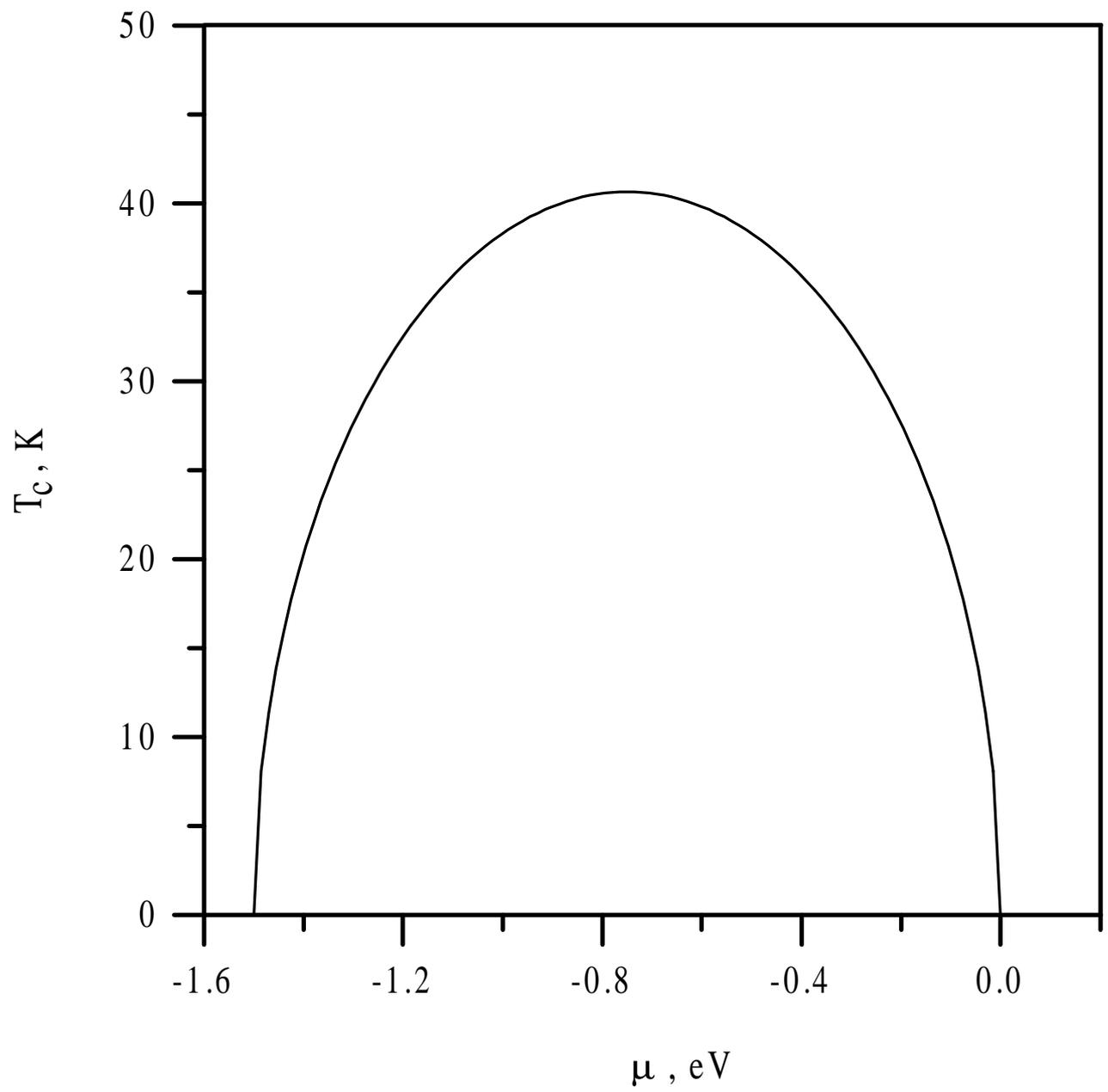
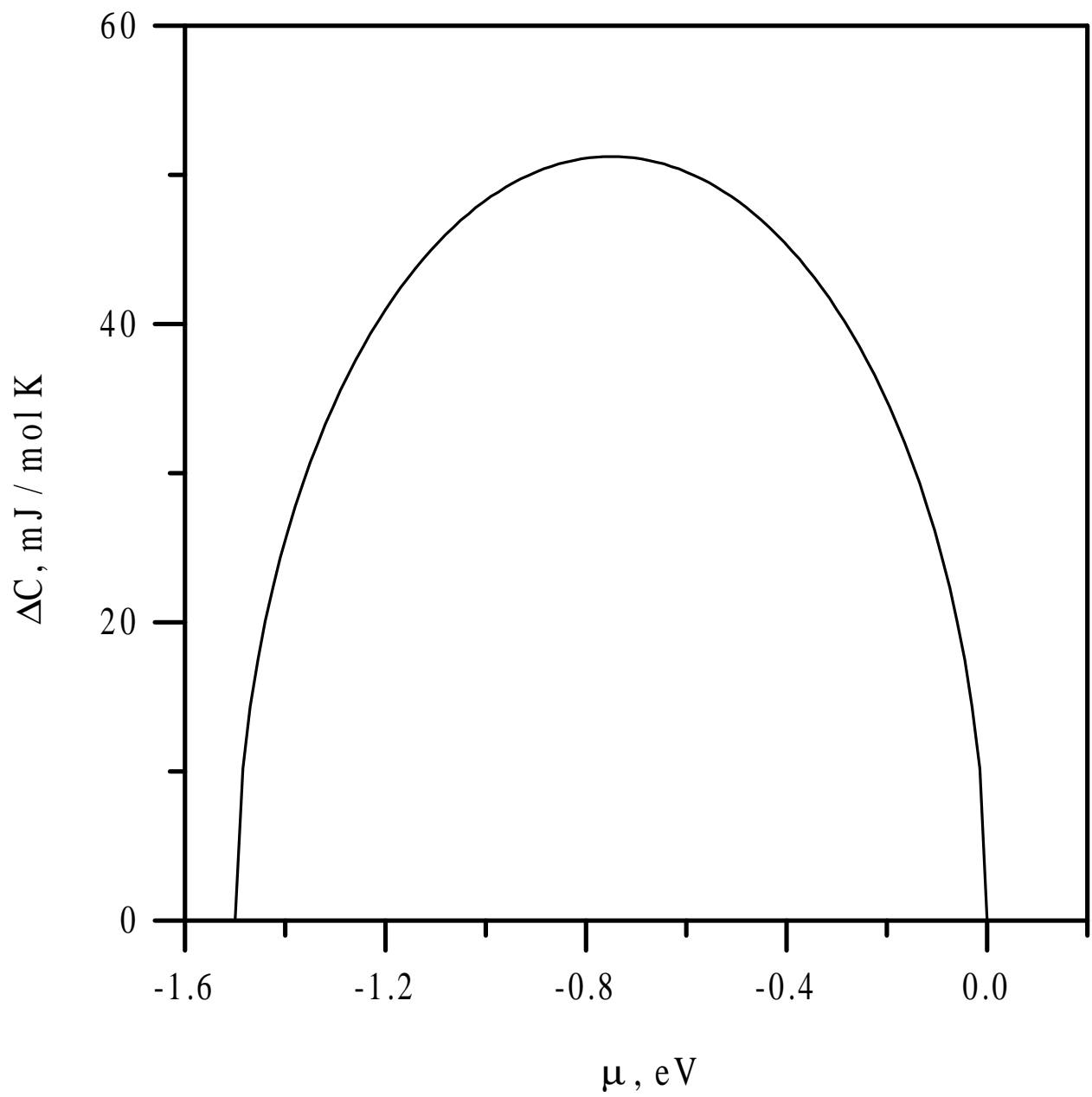
\end{document}